 \documentclass[prl,reprint,amssymb,twocolumn,noeprint,superscriptaddress,longbibliography]{revtex4-2}

 \usepackage[utf8]{inputenc}
\usepackage{amsmath}
\usepackage{graphicx} 
\usepackage{physics}
\usepackage{color}
\usepackage{amsfonts}
\usepackage{bm}
\usepackage{amssymb}
\usepackage{braket}
\usepackage{multirow}
\usepackage{epstopdf}
\usepackage[normalem]{ulem}

\usepackage[T1]{fontenc}

\usepackage[caption=false]{subfig}

\usepackage{hyperref}
\hypersetup{
colorlinks=true,
linkcolor=blue,
citecolor=blue,
filecolor=green,
urlcolor=blue,
}

\begin{document} 

\title{Highly spin-polarized molecules via collisional microwave pumping}

\author{Rebekah Hermsmeier}
\affiliation{Department of Physics, University of Nevada, Reno, Nevada, 89557, USA}
\author{Timur V. Tscherbul}
\affiliation{Department of Physics, University of Nevada, Reno, Nevada, 89557, USA}

\date\today

\begin{abstract}
We propose a general technique  to produce cold spin-polarized molecules in the electronic states of $\Sigma$ symmetry, in which rotationally excited levels are first populated by coherent microwave excitation, and then allowed to spin-flip and relax via collisional quenching, which populates a single final spin state. The steady-state spin polarization is maximized in the regime, where collisional slip-flipping transitions in the ground rotational manifold  ($N=0$)  are suppressed by a factor of $\geq$10 compared to those in the first rotationally excited  manifold ($N=1$), as generally expected for $\Sigma$-state molecules at temperatures below the rotational spacing between the $N=0$ and $N=1$ manifolds.
 We theoretically demonstrate the high selectivity of the technique for $^{13}$C$^{16}$O molecules immersed in a cold buffer gas of helium atoms, achieving a high degree  ($\geq$95\%) of nuclear spin polarization at 1~K.
 \end{abstract}
\maketitle

{\it Introduction}.
Spin-polarized quantum systems, such as atoms and molecules prepared  in single electron and/or nuclear spin states, are an essential resource in quantum science \cite{Nielsen:10,DiVincenzo:00}, high-precision spectroscopy \cite{Carr:09,Bohn:17,Patterson:13,Changala:19,Liu:22,Liu:23,Barskiy:17,Hirsch:15,Bucher:20}, and cold controlled chemistry \cite{Krems:08,Balakrishnan:16,Liu:22b}.
Indeed, the primary step in the vast majority of quantum algorithms and protocols involves the preparation of an ensemble of qubits initialized in a single pure quantum state \cite{Nielsen:10}, so the ability to initialize single quantum states with high fidelity is  an essential requirement for any quantum system considered as a potential qubit \cite{Nielsen:10,DiVincenzo:00}. Nuclear spin states are particularly advantageous for quantum information storage because of their weak coupling to the environment \cite{Omanakuttan:21,Gregory:21,Tscherbul:23}.  In spectroscopic experiments, single spin state initialization offers the benefits of enhanced spectral resolution and detection sensitivity, which are leveraged by, e.g.,  hyperpolarized  nuclear magnetic resonance (NMR) experiments \cite{Barskiy:17,Hirsch:15,Bucher:20,Walker:97,Gentile:17}.


Direct cooling methods offer a straightforward and efficient way to initialize molecules in single  quantum states \cite{Carr:09,Bohn:17}.
 In particular, buffer gas cooling can be used to create molecular samples with both the internal and external degrees of freedom cooled down to 1-100~K via momentum-transfer collisions with noble gas atoms \cite{Egorov:01,Lu:09,Patterson:13,Satterthwaite:23,Maussang:05,Singh:13,Iwata:17,Santamaria:21,Hofsass:21,Daniel:21,Changala:19,Liu:22,Liu:23}. This technique has enabled high-resolution spectroscopy of diatomic and polyatomic molecules \cite{Patterson:13,Iwata:17,Santamaria:21,Hofsass:21,Daniel:21,Changala:19,Liu:22,Liu:23}, enantiomer-specific detection of chiral molecules \cite{Patterson:13} and the creation of high-flux buffer-gas beams \cite{Maxwell:05,Patterson:09,Hutzler:12}.  However, high-fidelity initialization of individual spin states requires cooling to temperatures which are small compared to the Zeeman splitting (e.g, $T\leq10$~$\mu$K for the nuclear spin states of $^{13}$C$^{16}$O and magnetic fields $B\leq 0.1$~T).  These extremely low temperatures require advanced cooling methods (i.e., laser cooling, Stark deceleration, or  photoassociation) currently applicable only to a small (but growing) subset of molecules. This subset includes alkali-metal dimers KRb, NaK, RbCs, and NaCs \cite{Bohn:17},  laser-coolable molecular radicals  CaF, SrF, CaH, YO, and SrOH \cite{Shuman:10,Barry:14,Truppe:17,Anderegg:18,McCarron:18,Wu:21,Vilas:22} and molecules, which can be Stark or Zeeman decelerated in molecular beams, such as OH, NO, ND$_3$ \cite{Meerakker:06,Vogels:18,Tang:23}, CH$_3$, and O$_2$ \cite{Segev:19,Karpov:20,Liu:17,Cremers:18}. This limits the utility of extreme cooling as a general tool to achieve molecular spin polarization.


Besides direct cooling methods, a large number of experimental techniques have been developed for the production of spin-polarized species, which include optical pumping \cite{Drullinger:69,Zare:88,Happer:10book,Auzinsh:10book,Auzinsh:95book},  stimulated Raman adiabatic passage \cite{Vitanov:17}, photodissociation \cite{Rakitzis:03,Sofikitis:07,Spiliotis:21}, infrared excitation of molecular beams \cite{Rakitzis:05,Kannis:21,RubioLago:06}, and spin-selective chemical reactions   \cite{Barskiy:19}.  Specifically, in optical pumping, unidirectional population transfer is accomplished by a coherent drive followed by the action of a dissipative mechanism, which populates a set of target states via, e.g., spontaneous emission \cite{Drullinger:69,Zare:88,Happer:10book,Auzinsh:10book,Auzinsh:95book}, spin-exchange collisions \cite{Walker:97,Gentile:17} or  stimulated emission into a lossy cavity mode \cite{Wallquist:08}.
The diverse selection rules inherent to these dissipative transitions allow one to populate scientifically interesting final states, such as low-lying rovibrational levels of ultracold molecules  \cite{Manai:12,Viteau:08,Shuman:10,Barry:14,Truppe:17,Anderegg:18,McCarron:18,Wu:21,Vilas:22} and super-rotor states of  molecular ions \cite{Antonov:21}.


While standard optical pumping techniques based on spontaneous emission have found widespread use in molecular physics \cite{Zare:88,Auzinsh:95book}, their efficiency is limited by  the lack of selectivity of the emission process, which tends to populate a wide range of final rovibrational states \cite{Stuhl:08,Manai:12,Dragan:23} (an important exception are laser-coolable molecules, which have nearly diagonal matrices of Franck-Condon factors \cite{Shuman:10,Barry:14,Anderegg:18,McCarron:18,Wu:21,Truppe:17,Vilas:22}).
  As a result, conventional optical pumping cannot presently be used to produce spin-polarized molecules in single rovibrational and hyperfine states, which are crucial  for  hyperpolarized NMR,  high-precision spectroscopy, and quantum information science.

\begin{figure}
\captionsetup[subfigure]{margin={-0.9cm,0.5cm}}
    \centering
\subfloat[] {
  \includegraphics[width=0.65\columnwidth, trim = 0 150 15 30]{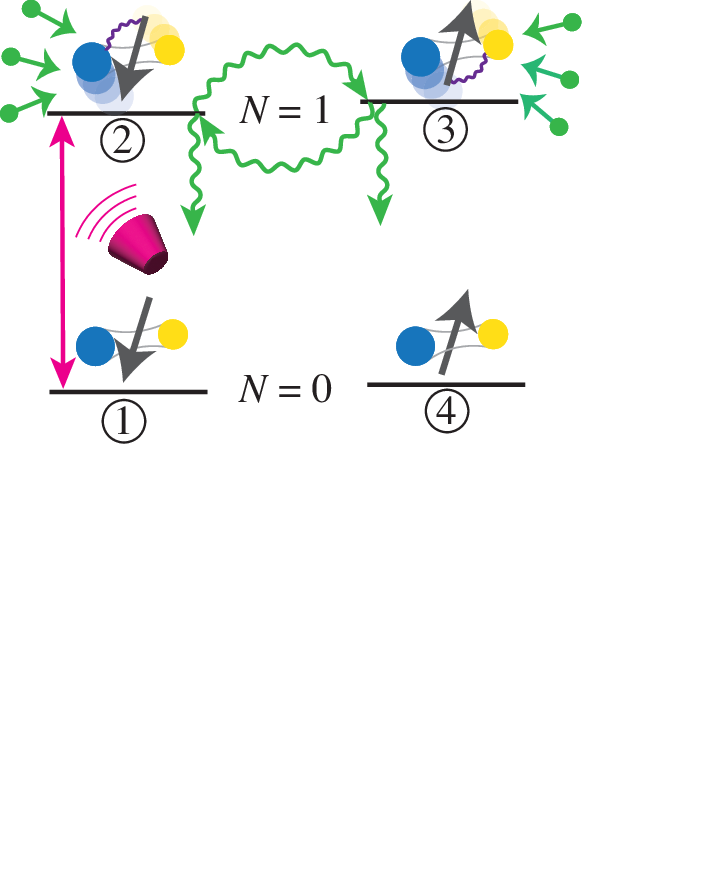}
}
\subfloat[]{
  \includegraphics[width=0.34\columnwidth,trim = 129 0 0 30]{fig_r_ratio.eps}
}
 \caption{(a) Schematic illustration of  collisional microwave pumping. We start with an incoherent mixture of rotationally ground state ($N = 0$) molecules in their spin-up and spin-down states $\ket{1}$ and $\ket{4}$. Resonant mw pumping (straight vertical arrow) populates the $N = 1$, spin-down state $\ket{2}$. Spin-flipping collisions with buffer gas atoms (wavy arrows) transfer population to the state $\ket{3}$, which undergoes collision-induced rotational relaxation down to the $N = 0$ spin-up state $\ket{4}$, where the molecules accumulate. (b)   Steady-state population ratio $r=\rho_{44}/(\rho_{11}+\rho_{22}+\rho_{33})$  [see Eq.~\eqref{efficiency_general}] as a function of reduced temperature $kT/\Delta E$ for different values of $W_{41}/W_{23}$ (see the legend).
  The ratio $W_{14}/W_{23}=6.9\times 10^{-9}$ corresponds to $^4$He~+~$^{13}$C$^{16}$O collisions at $T=1$~K and $B=0.05$~T \cite{Hermsmeier:23}. Higher values of $W_{14}/W_{23}$ are used to illustrate  the effect of collisional spin relaxation of the target state on the steady-state spin polarization. }
 \label{fig:model}
\end{figure}

Here, we overcome this limitation by proposing a microwave analog of optical pumping, which relies on   inelastic collisions to drive dissipative transitions to the desired final state. The only essential requirement is that of long relaxation times of the target spin sublevels, which is readily met for, e.g, nuclear spin states of $^{13}$C$^{16}$O($N=0$) \cite{Hermsmeier:23} and $N=0$ electron spin states of CaH and NH  \cite{Maussang:05,Singh:13,Krems:03,Krems:03b,Campbell:07,Campbell:09} in a cold $^4$He buffer gas.
  The idea of collisional microwave pumping  is illustrated in Fig.~1(a).
We start from an ensemble of molecules in their ground electronic and vibrational states, which are initially in a fully thermalized incoherent mixture of  spin-down and spin-up sublevels $\ket{1}$ and  $\ket{4}$ of the ground rotational state ($N=0$). The goal is to create a highly non-equilibrium spin distribution with most molecules in the spin-polarized state $\ket{4}=\ket{\uparrow}$.
 A resonant microwave drive causes rotational excitation  on the transition $|1\rangle \leftrightarrow |2\rangle$, which populates the $N=1$ rotational sublevel $\ket{2}$  with the same spin projection as the initial state $\ket{1}$. Cold collisions with  buffer-gas atoms cause spin-conserving rotational relaxation back to state $\ket{1}$, as well as, crucially,  spin-flipping  transitions to the spin-up sublevel $\ket{3}$ within the $N=1$ manifold (wavy lines in Fig. 1).
Once in state $\ket{3}$, molecules undergo rapid rotational relaxation  to the final spin-up level $\ket{4}$ in the $N=0$ manifold. If collisional spin relaxation in the $N=0$ manifold is strongly suppressed (see below), the molecular population is expected to accumulate in state $|4\rangle$ over time, leading to a steady growth of spin polarization to values as high as $95$\%  for molecules such as $^{13}$C$^{16}$O under realistic experimental conditions.

We note that the suppression of nuclear spin-flipping collisions in the $N=0$ manifold is generally expected for closed-shell molecules in non-degenerate electronic states of $\Sigma$ symmetry in the regime $kT< \Delta E$, where $\Delta E\simeq 2 B_e$ is the energy spacing between the levels $|1\rangle$ and $|2\rangle$, $B_e$ is the rotational constant and $k$ is the Boltzmann constant. This is because such collisions are mediated by intramolecular hyperfine interactions, which vanish identically for $N=0$  \cite{Hermsmeier:23}.
 This favorable situation extends to electron spin-flipping transitions in $\Sigma$-state radicals \cite{Maussang:05,Singh:13,Krems:03,Krems:03b,Campbell:07,Campbell:09,Tscherbul:11ch2} but not to rotationally excited states ($N\ge 1$) or to molecules in non-$\Sigma$ electronic states such as OH$(^2\Pi)$ and NO$(^2\Pi)$. For $^2\Pi$ molecules, the suppression can be engineered by superimposed electric and magnetic fields \cite{Tscherbul:09}.

Thus, we expect our scheme to be generally applicable to $\Sigma$-state molecules, which account for the vast majority of molecular species.


{\it Theoretical model.} 
We now present the general theory of collisional microwave pumping for polar molecules in cold buffer-gas environments before applying it  to an experimentally relevant case of $^{13}$C$^{16}$O($^1\Sigma^+$) molecules in a cold gas of He atoms.
A minimal model of collisional microwave pumping includes  4 molecular levels, as shown in Fig. 1(a). We use the density matrix formalism to describe  coherent mw pumping and collisional relaxation of molecular states on the same footing \cite{Blum}. 
The initial state of the molecules (before the  pumping is turned on) is described by the diagonal density matrix corresponding to the incoherent mixture of states $|1\rangle$ and $|4\rangle$, 
 ${\rho}(t=0)=\frac{1}{2}(|1\rangle\langle 1| + |4\rangle\langle 4|)$. The time evolution of the density matrix is governed by the optical Bloch equations \cite{CTbook}, which describe mw pumping  and  the Pauli rate equations, which describe collisional relaxation and decoherence. The resulting quantum master equation in the eigenstate basis takes the form \cite{Ramakrishna:05,Ramakrishna:06}
 \begin{align}\label{master_eq}\notag
\dot{\rho}_{11}(t) &=\frac{i\Omega}{2} (\rho_{12}-\rho_{21})+\sum_{n\neq1} \rho_{nn} (t)  W_{n1}-\rho_{11}(t)\sum_{n\neq1}  W_{1n}, \\ \notag
\dot{\rho}_{22}(t) &= \frac{i\Omega}{2} (\rho_{21}-\rho_{12})+ \sum_{n\neq2} \rho_{nn} (t)   W_{n2}-\rho_{22}(t) \sum_{n\neq2}  W_{2n}, \\ 
\dot{\rho}_{12}(t) &=-i\Delta \rho_{12}+\frac{i\Omega}{2}(\rho_{11}-\rho_{22}) -  \gamma_{12} \rho_{12},  \\ \notag
\dot{\rho}_{kk}(t) &= \sum_{n\neq k} \rho_{nn}(t)   W_{nk}-\rho_{kk} (t)\sum_{n\neq k}  W_{kn} \,\,(k=3,4),
\end{align} 
where $\rho_{mm}$ give the population of the $m$-th molecular state $(m=1{-}4)$, $\rho_{12}$ is the coherence between the  states $|1\rangle$ and $|2\rangle$,  $W_{nm}$ is the rate of the collision-induced  transition  $|n\rangle\to |m\rangle$,  $\Omega={E_0}\langle 1|\mathbf{d}\cdot \hat{\epsilon}|2\rangle/\hbar$ is the Rabi frequency for the mw transition $\ket{1}\leftrightarrow\ket{2}$, $\mathbf{E}(t)=E_0 \hat{\epsilon}\cos(\omega t)$ is the driving electric field with frequency $\omega$, amplitude $E_0$, and polarization vector $\hat{\epsilon}$, and $\mathbf{d}$ is the transition dipole moment.
In Eq.~\eqref{master_eq},  $\Delta=\hbar\omega - \Delta E$ is the mw detuning  and $\gamma_{12}$ is the dephasing rate. We assume resonant mw driving ($\Delta=0$).


 The steady-state solution of the master equation \eqref{master_eq} can be expressed 
 via the ratio of steady-state  populations   $r=\rho_{44}/(\rho_{11}+\rho_{22}+\rho_{33})$ \cite{SM}

  \begin{equation} \label{efficiency_general}
 r=\frac{1+ W_{41}/W_{23} +W_{41}/W_{21}}{3e^{-\Delta E/kT} + (2+ e^{-\Delta E/kT})W_{41}/W_{23} + 3W_{41}/W_{21}}.
 \end{equation}
The efficiency of collisional mw pumping is maximized when $r\gg 1$, i.e., when the target state population is large relative to those of all the other states. 
Equation~\eqref{efficiency_general} shows that  the ratio $W_{41}/W_{23}$  plays a crucial role in determining the steady-state spin polarization. In the limit where the ratios  $W_{41}/W_{23}$ and $W_{41}/W_{21}$  become small compared to $3 e^{-\Delta E/kT}$, 
 $r=\frac{1}{3}\exp(\frac{\Delta E}{kT})$  increases exponentially at low temperatures  [$kT/\Delta E\ll 1$, see Fig.~1(b)], leading to very high steady-state spin polarizations. Generally, one expects the following hierarchy of transition rates in cold collisions  of $\Sigma$-state molecules with structureless buffer gas atoms at $kT< \Delta E$: $W_{41} \ll W_{23} < W_{21}$ \cite{Hermsmeier:23}. The conditions $W_{41}/W_{23}\ll 1$ and $W_{41}/W_{21}\ll 1$ are thus expected to be valid for the vast majority of diatomic and small polyatomic molecules (e.g., for  $^{13}$C$^{16}$O  in $^4$He at 1~K, $W_{41}/W_{23} =6.9\times 10^{-9}$ and  $W_{41}/W_{21} =1.5\times 10^{-11}$ \cite{Hermsmeier:23}).
 
In the opposite limit of $W_{41}/W_{23}=1$ and $W_{41}/W_{21}\ll 1$, we obtain $r=(1+2e^{-\Delta E/kT})^{-1}$, which gives $r=1$ ($\Delta E/kT \gg 1$) and $r=1/3$  ($\Delta E/kT \ll 1$). In either case,  no  significant spin polarization is generated as expected due to the rapid thermalization of the $N=0$ spin sublevels via buffer-gas collisions.
In Fig. 1(b), we plot the steady-state spin polarization as a function of $\Delta E/kT$ for several values of $W_{41}/W_{23}$. We observe that while the desirable exponential trend $r=\frac{1}{3}\exp(\frac{\Delta E}{kT})$ holds for $W_{41}/W_{23}\leq 6.9\times 10^{-7}$ the steady-state polarization is significantly diminished for $W_{41}/W_{23} \geq 0.1$.
Thus, useful non-equilibrium spin polarizations ($r>1$) can only be achieved in the regime, where  $W_{41}/W_{23}\lesssim 0.1$.
\color{black}


 

\begin{figure}[t]
 \centering
 \includegraphics[width=8cm]{fig_COenergy_levels.eps}
 \caption{Rotational-Zeeman energy levels and transition frequencies of $^{13}$C$^{16}$O as a function of magnetic field. The corresponding energy levels of the four-level model are shown on the right. Each level is labeled by the quantum numbers $N$, $M_N$, and $M_I$ in the high-field limit.  }
  \label{fig:energyLevel}
\end{figure}


To explore the feasibility of collisional microwave pumping beyond the simple four-level model, consider a typical  closed-shell $^1\Sigma$ molecule  such as $^{13}$C$^{16}$O, immersed in a cold buffer gas of $^4$He atoms, as realized in a number of experiments \cite{Egorov:01,Lu:09,Patterson:13,Satterthwaite:23,Maussang:05,Singh:13,Iwata:17,Santamaria:21,Hofsass:21,Daniel:21,Changala:19,Liu:22,Liu:23}. The  $^{13}$C$^{16}$O molecule has a single magnetic nucleus with the nuclear spin $I=1/2$.
Figure~2 displays the Zeeman levels of $^{13}$C$^{16}$O obtained by diagonalization of the molecular Hamiltonian   \cite{Brown:03,Meerts:77,Klapper:00,Hermsmeier:23}. The levels are arranged in two rotational manifolds, with two levels in the ground ($N=0$) manifold  and six levels in the first rotationally excited ($N=1$) manifold. 
  We assume that the external magnetic field is directed along the space-fixed $z$-axis, and that  the Zeeman interaction is large 

    compared to the intramolecular hyperfine interaction in the $N=1$ manifold. This is necessary to decouple the nuclear spin degree of freedom from rotational motion,    enforcing the radiative and collisional selection rules (such as $W_{41} \ll W_{23} < W_{21}$) which are crucial for the optimal performance of collisional mw pumping. This requires $g_I \mu_N B\gg A$,  where $A$ is the hyperfine constant, $g_I$ is the nuclear $g$-factor, and $\mu_N$  is the nuclear Bohr magneton.
Typically, $A/2\pi\lesssim 1$~MHz, so the requisite $B$-fields  will rarely exceed 0.1 T. 
    \color{black}
The molecular eigenstates can be written as $|NM_N\rangle|IM_I\rangle=|NM_N,M_I\rangle$, where $|NM_N\rangle$ are the eigenstates of the squared rotational angular momentum operator $\hat{N}^2$ and its projection $\hat{N}_z$ on the $z$ axis defined by the $B$ field, and $|I M_I\rangle$ are the eigenstates of the squared nuclear spin angular momentum  $\hat{I}^2$ and its projection $\hat{I}_z$.

  We now identify the nuclear spin levels $|00, -1/2\rangle$  and $|11, -1/2\rangle$ of $^{13}$C$^{16}$O  with the states $|1\rangle$ and $|2\rangle$ of the four-level model (see Fig.~2). We further identify the levels $|10,1/2\rangle$ and $|00, 1/2\rangle$ with the states $|3\rangle$ and $|4\rangle$ of the four-level model. 
Because the electric transition dipole moment is independent of the nuclear spin, mw transitions can only be driven between the states of the same $M_I$, e.g., from the initial  state $|00,-1/2\rangle$ to the final states $|1 M_N,-1/2\rangle$ with $M_N=0,\pm1$.
The transition selected for mw pumping (here, $|00,-1/2\rangle\to |11,-1/2\rangle$) must have a unique frequency to prevent the simultaneous driving of multiple transitions (such as $|3\rangle\leftrightarrow |4\rangle$) causing undesirable population transfer.
For example, the difference in the energy gaps for the $|00,-1/2\rangle \leftrightarrow |10,-1/2\rangle$ and $|00,1/2\rangle \leftrightarrow |10,1/2\rangle$ transitions in $^{13}$C$^{16}$O is just $4\times 10^{-8}$~cm$^{-1}$ at $B=0.05$~T and hence these transitions will likely be driven simultaneously by a single mw frequency at a typical value of $\Omega = 10$~kHz.  We therefore choose the $|00,-1/2\rangle \leftrightarrow |11,-1/2\rangle$  transition for our mw pumping setup, since it is  detuned from the competing transition $|00,1/2\rangle \leftrightarrow |11, 1/2\rangle$ by $10^{-6}$~cm$^{-1}$.
We note that because molecular Zeeman levels  can be tuned by  an external magnetic field (see Fig.~2), finding a unique transition frequency should generally be possible unless the molecular spectrum is extremely dense.


Having specified the structure of the four-level model, we now explore the time dynamics of collisional microwave pumping. To this end, we solve Eqs.~\eqref{master_eq} numerically using accurate $^4$He~+~$^{13}$C$^{16}$O collision rates obtained from rigorous quantum scattering calculations based on an {\it ab initio} potential energy surface \cite{Hermsmeier:23,SM}.  
Figure 3(a) shows the time evolution of state populations in the four-level model of $^{13}$C$^{16}$O in a cold $^4$He buffer gas  ($T=1$~K) following the initial turn-on of the mw drive at $t=0$.
 At short timescales, we observe damped Rabi oscillations between the states $|00,-1/2\rangle$ and $|11,-1/2\rangle$ (or $|1\rangle$ and $|2\rangle$) involved in mw pumping. By setting the collision rates to zero and recovering the standard Rabi cycles, we verified that the damping is due to collisional relaxation.
   At later times ($t>0.4$~s) the population of  the target state $|4\rangle$ approaches 98\%, closely following the steady-state analytic solution of the 4-level model  \cite{SM} (98.5\%).
    These results  theoretically demonstrate that  collisional microwave pumping can be used to produce highly spin-polarized molecular ensembles.
    
  

To quantify the potentially detrimental effects of population leakage to the molecular states not included in the four-state model, we solved an extended quantum master equation, taking into account all 8 lowest rotational-Zeeman states of  $^{13}$C$^{16}$O [see Fig.~2].
The calculated population dynamics shown in Fig.~3(b) are closely similar to the four-level model results  [see Fig.~3(a)] and the steady-state efficiency (95.8\%) is only slightly  lower than that predicted by $r=\frac{1}{3}\exp(\Delta E/kT)$, demonstrating that population leakage does not significantly diminish the high value of steady-state spin polarization.

Collisional microwave pumping could be realized experimentally using large cryogenic buffer-gas cells used in state-of-the-art mw spectroscopy experiments  \cite{Porterfield:19} ($n\simeq 10^{14}$ cm$^{-3}$, $\Omega/2\pi=30$ kHz, $T=1$~K,  $B=0.05$~T, see \cite{SM} for state-to-state collision rates $W_{ij}$). The main limitations of the technique are related to the maximum buffer-gas density and the ratio of $N=0$ to $N=1$ spin flipping rates  $W_{41}/W_{23}$.
The pressure broadening parameter for the $N=0\to 1$ transition of CO in $^4$He was measured to be $80.437$ kHz/mTorr at $T=1.249$~K \cite{Beaky:96}, so 
the $1\leftrightarrow 2$ transition linewidth (80~kHz at $n=10^{16}$ cm$^{-3}$) can exceed the energy gap between the neighboring $N=1$ Zeeman levels, making it challenging to address specific mw transitions. Second, when $W_{41}/W_{23}\geq 0.1$, the steady-state spin polarization is rapidly destroyed by spin-flipping collisions with buffer-gas atoms.
 Fortunately, as argued above, these limitations do not play a significant role for $\Sigma$-state molecules at typical buffer-gas densities and low enough temperatures,  making it possible to achieve high steady-state spin polarizations [$r\gg 10$, see Fig.~1(b)].
 

\color{black}




\begin{figure}[t!]
 \centering
 \includegraphics[width=0.7\columnwidth, trim = 0 0 0 -10]{fig_time_dyn.eps}
 \caption{Molecular state populations as a function of time obtained by numerically solving the master equation \eqref{master_eq} for the four-level model (a) and the 8-level model of $^{13}$C$^{16}$O in a buffer gas of $^4$He (b) under realistic experimental conditions \cite{Egorov:01,Lu:09,Patterson:13,Satterthwaite:23,Maussang:05,Singh:13,Iwata:17,Santamaria:21,Hofsass:21,Daniel:21,Changala:19,Liu:22,Liu:23} (see text). 
 Mw pumping is turned on at $t=0$, when the molecules are initially in an equal  incoherent mixture of their $N=0$ nuclear spin states $\ket{1}$ and $\ket{4}$.  98\% (95.8\%) of  molecules end up in the target state after 0.4 s according to the four (8)-level model. }
  \label{fig:4Level}
\end{figure}

In summary, we have proposed a microwave analog of optical pumping, which can be used to drive  cold polar $\Sigma$-state molecules into single spin-polarized quantum states with high efficiency. The method relies on mw transitions to populate rotationally excited states and on inelastic collisions with buffer-gas atoms to drive relaxation to the final (target) states. We theoretically demonstrate the efficiency of collisional microwave pumping using an analytic four-state model, as well as a complete 8-state model for $^{13}$C$^{16}$O molecules in a cold $^4$He buffer gas. Our calculations predict spin-polarization efficiencies exceeding 95\% under realistic  experimental conditions.

Our method relies only on well-characterized rotationally excited molecular states and buffer gas collisions, and can therefore be extended to a large number of molecules, including those  which cannot be laser cooled.  It may also be extendable to polar molecules trapped in inert cryogenic matrix environments \cite{Fajardo:09,Fajardo:13}, which are promising candidates for precision spectroscopy and searches of new physics \cite{Kozlov:06,Vutha:18,Vutha:18b,Li:23,Rollings:23}.

 Collisional microwave pumping is conceptually similar to spin-exchange optical pumping \cite{Walker:97,Gentile:17}: In both methods, spin-changing collisions are used to create a non-equilibrium spin population. However, 
 our method is uniquely suited for molecules as it relies on {\it rotational} degrees of freedom 
 and inelastic collisions with commonly available buffer-gas atoms. As such, collisional microwave pumping does not require going beyond current buffer-gas cooling technology \cite{Egorov:01,Lu:09,Patterson:13,Satterthwaite:23,Maussang:05,Singh:13,Iwata:17,Santamaria:21,Hofsass:21,Daniel:21,Changala:19,Liu:22,Liu:23}, and may thus  be readily realizable in near-term experiments.

We thank Jonathan Weinstein and David Patterson for several   truly essential discussions in the early stages of this work. 
We gratefully acknowledge support from the NSF CAREER program (grant No. PHY-2045681).



%

\end{document}